\documentstyle[aps,psfig,tighten,floats]{revtex}

\newcommand{\SC}{\scriptscriptstyle} 
\newcommand{\SS}{\scriptstyle} 

\begin{document}

\title{\bf Spin foam model for Lorentzian General Relativity}

\author{Alejandro Perez and Carlo Rovelli \\ {\it Centre de 
Physique
Th\'eorique - CNRS, Case  907, Luminy,
             F-13288 Marseille, France}, and \\
{\it Physics Department, University of Pittsburgh, 
             Pittsburgh, Pa 15260, USA}}

\maketitle

\begin{abstract}

We present a spin foam formulation of Lorentzian quantum General
Relativity.  The theory is based on a simple generalization of an
Euclidean model defined in terms of a field theory over a group.  Its
vertex amplitude turns out to be the one recently introduced by
Barrett and Crane.  As in the case of its Euclidean relatives, the
model fully implements the desired sum over 2-complexes which encodes
the local degrees of freedom of the theory.

\end{abstract}
 
\section{Introduction}

Spin foam models provide a well defined framework for background
independent diffeomorphism invariant quantum field theory.  A
surprising great deal of approaches have led to this type of
models\cite{Reisenberger,Iwasaki,Baez,rr,Roberto,BC}.  In particular,
due to their non perturbative features, spin foam models appear as a
very attractive framework for quantum gravity.

Spin foam models provide a rigorous implementation of the
Wheeler-Misner-Hawking\cite{misner,haw} sum over geometries
formulation of quantum gravity.  The 4-geometries summed over are
represented by foam-like structures known as spin foams.  They are
defined as colored 2-complexes.  A 2-complex $J$ is a (combinatorial)
set of elements called ``vertices'' $v$, ``edges'' $e$ and ``faces''
$f$, and a boundary relation among these, such that an edge is bounded
by two vertices, and a face is bounded by a cyclic sequence of
contiguous edges (edges sharing a vertex).  A spin foam is a 2-complex
plus a ``coloring'' $N$, that is an assignment of an irreducible
representation $N_{f}$ of a given group $G$ to each face $f$ and of an
intertwiner $i_{e}$ to each edge $e$.  The model is defined by the
partition function
\begin{equation}
Z = \sum_{J}\ {\cal N}(J) \sum_{N}\ \prod_{f\in J} 
A_{f}(N_{f}) \ \prod_{e\in J} A_{e}(N_e) \prod_{v\in
J} A_v(N_v),  \label{Z}
\end{equation}
where $A_{f}$, $A_{e}$ and  $A_v$ correspond to the amplitude associated 
to faces, edges,  and vertices respectively (they are given functions of the 
corresponding colors). ${\cal N}(J)$ is a normalization factor
for each 2-complex. 

Spin foam models related to gravity have been obtained as modifications
of topological quantum field theories (corresponding to BF theory) by
implementation of the constraints that reduce BF theory to general
relativity\cite{BC,Reisenberg97,iwa0,ac}.  So far, these constructions
were restricted to the Euclidean sector.  A crucial step towards the
definition of a physical Lorentzian model has been taken by Barrett
and Crane in \cite{BC2}.  In this work, Barrett and Crane construct a
well defined vertex amplitude for Lorentzian quantum gravity, based on
the representation theory of $SL(2,C)$.

Based on the work of Barrett and Crane, in this letter we complete the
definition of a Lorentzian spin foam model for gravity.  That is, we
give an explicitly formula for the partition function of the model. 
To this aim, we use the technology provided by field theory over group
manifolds, developed in in \cite{dfkr,cm}.  In this language, spin
foams (quantum 4-geometries) appear as the Feynman diagrams of a
certain nonlocal scalar field theory over a group.  Strikingly, the
Barrett-Crane Lorentzian vertex appears completely naturally in this
context.

Two important points should be emphasized.  First, the theory defined
in this way implements automatically the sum over 2-complexes $J$ in
(\ref{Z}), and in particular, fixes the ${\cal N}(J)$ value.  
This sum is necessary to restore full general covariance
of a theory with local degrees of freedom such as GR \cite{Baez,dfkr}. 
Indeed, in the case of a topological field theory
\cite{t,Ooguri:1992b,CraneYetter,CraneYetter1} the sum over
2-complexes in (\ref{Z}) can be dropped (for fixed topology) due to
the triangulation invariance of the partition function.  This is a
consequence of the absence of local degrees of freedom in the
topological theory.  When the constraints are implemented, however,
the theory acquires the local degrees of freedom of gravity and
different 2-complexes carry physical information.  In the language of
standard QFT, they represent higher order radiative corrections.  In
our model, the sum over 2-complexes is automatically implemented by
the formalism. 

The second point is about divergences.  The Euclidean model in
\cite{BC} is defined in terms of a quantum deformation of the gauge
group ($SO_q(4)$, with $q^n =1$).  The quantum deformation is needed
to regularize divergences in (\ref{Z}).  In the limit in which the
quantum deformation is removed ($q \rightarrow 1$), these divergences
appear whenever the 2-complex $J$ includes bubbles \cite{ac}.  In
reference \cite{ac}, using the field theory over group technology, we
have defined a variant of the model, in which the basic bubble
amplitudes are finite for $q=1$.  The definition of the Lorentzian
model presented here corresponds to this variant.  Although further
study is certainly needed, we suspect that the Lorentzian model
presented here might be finite even with $q=1$.

Many issues remain open.  In particular: (i) Can we get stronger
evidence that the model gives general relativity in the classical
limit?  (ii) Can finiteness be proven?  (iii) What is the physical
meaning and the physical regime of validity of the expansion in the
number of vertices?  (iv) Do the transition amplitudes of the model
have a direct physical interpretation?  If answers to these questions
turned out to be positive, the model presented here might provide an
interesting candidate for a quantum theory of gravity.

In the bulk of the paper we introduce the new model and discuss its
properties.  In an appendix we present a compendium of known results
on harmonic analysis and representation theory of $SL(2,C)$ on which
our construction is based.

\section{$SL(2,C)$ state sum model of Lorentzian QG}

\begin{figure}[h]
\centerline{{\psfig{figure=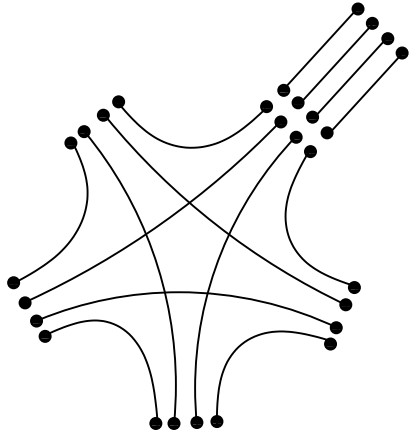,height=6cm}}}
\label{vertex}\bigskip \caption{Structure of the interaction.  The
black circle represent the projections $\delta_{0j}$ (\ref{proj})
produced by the $SU(2)$ integrations in (\ref{mart}).}
\label{figure}
\end{figure}

We start with a field $\phi(g_1,g_2,g_3,g_4)$ over 
${ SL(2,C)\times SL(2,C)\times SL(2,C)\times SL(2,C)}$.  We
assume the field has compact support and is symmetric under arbitrary
permutations of its arguments\footnote{ This symmetry guarantees
arbitrary 2-complexes $J$ to appear in the Feynman expansion\cite{dfkr}.}.  We define the projectors $P_{\gamma}$ and
$P_{u}$ as
\begin{equation} 
    \label{pg} P_{g}\phi(g_i)
\equiv \int d\gamma  \ {\phi}(g_i\gamma),
\end{equation} 
and
\begin{equation} \label{pu}        
P_{u}\phi(g_i) \equiv \int du_i \ {\phi}( g_i u_{i}),
\end{equation}
where $\gamma \in SL(2,C)$, and $u_i \in SU(2)$, and $d\gamma$, $du$
denote the corresponding invariant measures.  We define the action of
our model as 
\begin{equation}
    \label{action} 
S[\phi]=\int dg_i \left[ P_{\gamma} \phi(g_i) \right]^2 + {\lambda
\over 5!} \int dg_i \left[ P_{\gamma} P_{u}\phi(g_i) \right]^5,
\end{equation} 
where $\gamma_{i} \in SL(2,C)$, $\phi(g_i)$ denotes
$\phi(g_1,g_2,g_3,g_4)$, and the fifth power in the interaction term
is notation for 
\begin{equation}
 \left[\phi(g_i)\right]^5:=\phi(g_1,g_2,g_3,g_4)\
 \phi(g_4,g_5,g_6,g_7)\ \phi(g_7,g_3,g_8,g_9)\
 \phi(g_9,g_6,g_2,g_{10}) \ \phi(g_{10},g_8,g_5,g_1).
\end{equation}
The $\gamma$ integration projects the field into the space of gauge
invariant fields, namely, those such that $\phi(g_i)=\phi(g_i\mu)$ for
$\mu \in SL(2,C)$.%
\footnote{Because of this gauge invariance, the action (\ref{action})
is proportional to the trivial diverging factor $\int d\gamma$.  This
divergence could be fixed easily, for instance by gauge fixing and
just dropping one of the group integrations.  For the clarity of the
presentation, however, we have preferred to keep gauge invariance
manifest, use the action formally to generate the Feynman expansion,
and drop the redundant group integration whenever needed.\label{foot}}
The vertex and propagator of the theory are simply given by a set of
delta functions on the group, as illustrated in \cite{ac}, to which we
refer for details.  Feynman diagrams correspond to arbitrary 2-complex
$J$ with 4-valent edges (bounding four faces), and 5-valent vertices
(bounding five edges).  Once the configuration variables $g_i$ are
integrated over, the Feynman amplitudes reduce to integrals over the
group variables $\gamma$ and $u$ in the proyectors in (\ref{action}). 
These end up combined as arguments of one delta functions per face
\cite{ac}.  That is, a straightforward computation yields
\begin{equation} 
    \label{mart}A(J)=\int du d\gamma \prod_{{e }}
\prod_{{f }} \delta(\gamma^{\SC(1)}_{e_{1}}u^{\SC }_{\SC
1f}\gamma^{\SC(2)}_{e_{1}}u^{\SC \prime }_{\SC
1f}\gamma^{\SC(3)}_{e_{1}} \dots \gamma^{\SC(1)}_{e_{N}}u^{\SC }_{\SC
Nf}\gamma^{\SC(2)}_{e_{N}}u^{\SC \prime }_{\SC
Nf}\gamma^{\SC(3)}_{e_{N}}). 
\end{equation} 
In this equation, $\gamma^{\SC(1)}_{e_{}}$, and
$\gamma^{\SC(3)}_{e_{}}$ come from the group integration in the
projectors $P_{\gamma}$ in the two vertices bounding the edge $e$. 
$\gamma^{\SC(2)}_{e_{}}$ comes from the projector $P_{\gamma}$ in the
propagator defining the edge $e$.  Finally, $u^{\SC }_{\SC 1f}$ and
$u^{\SC \prime }_{\SC 1f}$ are the $SU(2)$ integration variables in
the projector $P_{h}$ in the two vertices.  Notice that each $u$
integration variable appear only once in the integrand, while each
$\gamma$ integration variable appears in four different delta's (each
edge bounds four faces).  The index $N$ denotes the number of edges of
the corresponding face.  Now we use equation (\ref{vani}) to expand
the delta functions in terms of irreducible representations of
$SL(2,C)$. Only the representations $(n,\rho)$ in the principal series
contribute to this expansion. We obtain 
\begin{equation}
    \label{aj}
A(J)=\sum_{n_f} \int_{\rho_f} d\rho_{f} \prod_{{f }} (\rho_f^2+n_f^2) \int
\prod_{{e }} d\gamma du \ {\rm Tr}\left[ T_{n_f
\rho_f}(\gamma^{\SC(1)}_{e_{1}}u^{\SC }_{\SC
1f}\gamma^{\SC(2)}_{e_{1}}u^{\SC \prime }_{\SC
1f}\gamma^{\SC(3)}_{e_{1}} \dots \gamma^{\SC(1)}_{e_{N}}u^{\SC }_{\SC
Nf}\gamma^{\SC(2)}_{e_{N}}u^{\SC \prime }_{\SC
Nf}\gamma^{\SC(3)}_{e_{N}} )\right].  
\end{equation} 
Next, we rewrite this equation in terms of the matrix elemets 
$D^{n \rho}_{j_1 q_1 j_2 q_2}(\gamma)$ of the representation $(n,\rho)$ 
in the canonical basis, defined in the
appendix.  The trace becomes
\begin{eqnarray}
\nonumber && {\rm Tr}\left[
T_{n_f \rho_f}(\gamma^{\SC(1)}_{e_{1}}u^{\SC }_{\SC
1f}\gamma^{\SC(2)}_{e_{1}}u^{\SC \prime }_{\SC 1f}\gamma^{\SC(3)}_{e_{1}}
\dots \gamma^{\SC(1)}_{e_{N}}u^{\SC }_{\SC Nf}\gamma^{\SC(2)}_{e_{N}}u^{\SC
\prime }_{\SC Nf}\gamma^{\SC(3)}_{e_{N}}  )\right] =  \\ 
&& D^{n_f \rho_f}_{j_1 q_1 j_2 q_2}(\gamma^{\SC(1)}_{e_{1}}) 
D^{n_f \rho_f}_{j_2 q_2 j_3 q_3}(u^{\SC }_{\SC1f}) 
D^{n_f \rho_f}_{j_3 q_3 j_4 q_4}(\gamma^{\SC(2)}_{e_{1}}) 
\dots D^{n_f \rho_f}_{j_{.} q_{.} j_1 q_1}(\gamma^{\SC(3)}_{e_{N}}). 
\end{eqnarray}
(Repeated indices are summed, and the range of the $j_{n}$ and $q_{n}$
indices is specified in the appendix.)  According to equation
(\ref{proj}), each $u$ integration produces a projection into the
subspace spanned by the simple representations $(0,\rho)$.%
\footnote{This projection implements the constraint that reduces BF
theory to GR. Indeed, the generators of $SL(2,C)$ are identified with
the classical two-form field $B$ of BF theory.  The generators of the
simple representations satisfy precisely the BF to GR constraint. 
Namely $B$ has the appropriate $e\wedge e$ form \cite{BC,Baez}. 
Notice however that the representations $(0,\rho)$ are not the only
simple representations; there are also simple representations of the
form $(n,0)$ with $n=1,2\dots$.  The two sets have a simple
geometrical interpretation in terms of space and time like directions
(see \cite{BC2}).  We suspect that to recover full GR both set of
simple representations should be included.}
That is, after the integration over $u^{\SC }_{\SC1f}$, the matrix
$D^{n_f \rho_f}(u^{\SC }_{\SC1f})_{j_2 q_2 j_3 q_3}$ collapses to
$\delta_{j_1 0}\delta_{j_2 0}$.  One of these two Kroeneker deltas
appears always contracted with the indices of the $D(\gamma)$
associated to a vertex; while the other is contracted with a
propagator.  We observe that the representation matrices
associated to propagators ($\gamma^{(2)}_{e}$) appear in four faces in
(\ref{aj}).  The ones associated to vertices appear also four times
but combined in the ten corresponding faces converging at a vertex. 
Consequently, they can be paired according to the rule $D^{n \rho}_{j
q k l}(\gamma_{e_i})D^{n \rho}_{k l s t}(\gamma_{e_j}) =D^{n
\rho}(\gamma_{e_i} \gamma_{e_j})_{j q s t}$.  In Fig.\,(\ref{figure})
we represent the structure described above.  A continuous line
represents a representation matrix, while a dark dot a contraction
with a projector ($\delta_{j 0}$).  Taking all this into account, we 
have
\begin{eqnarray}
    \label{lsf} 
    A(J)=\sum_{n_f} \int_{\rho_f} \prod_{{f }}\ (\rho_f^2+n_f^2) \ \
    \prod_{{e }} \ A_e( \rho_{\SC e_1},\dots \rho_{\SC e_{4}};n_{\SC
    e_1},\dots n_{\SC e_{4}} ) \ \prod_{{ v }} \ A_v(\rho_{\SC v_1},
    \dots \rho_{\SC v_{10}};n_{\SC v_1}, \dots n_{\SC v_{10}}),
    \end{eqnarray} 
where $A_e$ is given by
\begin{equation}\label{edge}
A_e( \rho_{\SC e_1},\dots \rho_{\SC e_{4}};n_{\SC e_1},\dots n_{\SC
e_{4}})= \delta_{n_{\SC e_1 0}} \dots \delta_{n_{\SC e_4 0}} \int
d\gamma \ D^{0 \rho_1}_{0000}{\SS (\gamma)} \dots D^{0
\rho_4}_{0000}{\SS (\gamma)},
\end{equation}
and $A_{v}$ by
\begin{eqnarray}
\label{choclo}\nonumber && A_v(\rho_{\SC
v_1}, \dots \rho_{\SC v_{10}};n_{\SC
v_1}, \dots n_{\SC v_{10}}) = \delta_{n_{\SC v_1 0}} 
\dots \delta_{n_{\SC v_{10} 0}}  \\ \nonumber &&
\int \  \prod^{5}_{i=1} d\gamma_i \ 
D^{\SC 0 \rho_1}_{\SC 0000}{\SS (\gamma_1\gamma^{-1}_5)} 
D^{\SC 0 \rho_2}_{\SC 0000}{\SS (\gamma_1\gamma^{-1}_4)}  
D^{\SC 0 \rho_3}_{\SC 0000}{\SS (\gamma_1\gamma^{-1}_3)} 
D^{\SC 0 \rho_4}_{\SC 0000}{\SS (\gamma_1\gamma^{-1}_2)} \\ &&
D^{\SC 0 \rho_5}_{\SC 0000}{\SS (\gamma_2\gamma^{-1}_5)} 
D^{\SC 0 \rho_6}_{\SC 0000}{\SS (\gamma_2\gamma^{-1}_4)} 
D^{\SC 0 \rho_7}_{\SC 0000}{\SS (\gamma_2\gamma^{-1}_3)} 
D^{\SC 0 \rho_8}_{\SC 0000}{\SS (\gamma_3\gamma^{-1}_5)} 
D^{\SC 0 \rho_9}_{\SC 0000}{\SS (\gamma_3\gamma^{-1}_4)} 
D^{\SC 0 \rho_{10}}_{\SC 0000}{\SS (\gamma_4\gamma^{-1}_5)} .  
\end{eqnarray} 
In Fig.\,(\ref{figure}), each $D^{0 \rho_1}_{0000}(\gamma)$ in
the previous expressions corresponds to a line bounded by two dark dots.

The functions $D^{0 \rho_1}_{0000}(\gamma)$ are known explicitly in
the literature\cite{ru}; they can be realized as
functions on the hyperboloid ($H^{+}$) $x^{\mu} x_{\mu}=1$, $x_0>0$ in
Minkowski spacetime in the following way.  Any $\gamma \in SL(2,C)$
can be written as $\gamma=u_1 d\, u_2$ with $u_i \in SU(2)$ and
\begin{equation}
d=\left[\begin{array}{c}e^{\eta/2} \ \ 0 \\ 0 \ \ e^{-\eta/2}
\end{array} \right].  
\end{equation} 
(Any Lorentz transformation can be obtained with a rotation, a boost
in the $z$ direction and another rotation.)  In this
parametrization, $D^{0 \rho}_{0000}(\gamma)$ is a function of $\eta$
only.  We denote it as $K(\eta)$.  Its form is given in \cite{ru} (page 65) as
\begin{equation}
K_{\rho}(\eta)={2\; {\rm sin}(1/2 \eta \rho) \over \rho \; {\rm
sinh}(\eta)}.
\end{equation} 
Given $\gamma\in SL(2,C)$ then $x_{\gamma}:=\gamma \gamma^{\dagger}$
represents a point in $H^{+}$.  It is easy to see that the parameter
$\eta$ associated to $\gamma$ corresponds to the hyperbolic distance
from the point $x_{\gamma}$ to the apex of the hyperboloid (boost
parameter).  The hyperboloid is a transitive surface under the action
of $SL(2,C)$, i.e., it is Lorentz invariant.  Therefore, the parameter
$\eta$ associated to a product $\gamma_1\gamma^{-1}_2 \in SL(2,C)$
corresponds to the hyperbolic distance of the point 
$\gamma_{2}^{-1}[x_{\gamma_1}]$ to the apex \footnote{We denote by
$\gamma[x]$ the usual action of $SL(2,C)$ matrices on $x$ defined as
an hermitian spinor, namely, $\gamma[x]=\gamma x \gamma^{\dagger}$.}. 
Equivalently, it corresponds to the hyperbolic distance between
$x_{\gamma_1}$ and the Lorentz transformed apex 
$\gamma_2\gamma^{\dagger}_2:=x_{\gamma_2}$(namely,
$\eta(\gamma_1\gamma^{-1}_2) ={\cal D}_{\SC H^{+}}(x_{\gamma_1},
x_{\gamma_{2}})$ ).  We define 
\begin{equation} 
D^{0 \rho}_{0000}(\gamma_1\gamma^{-1}_2
)=K_{\rho}(\eta(\gamma_1\gamma^{-1}_2)):=K_{\rho}(x_1, x_2). 
\end{equation}
Finally, the invariant measure on $SL(2,C)$ is simply the product of
the invariant measures of the hyperboloid and $SU(2)$, that is
$d\gamma=du\, dx$.  Using all this, the vertex and edge amplitudes can
be expressed in simple form.  The edge amplitude (\ref{edge}) becomes 
\begin{equation}
    \label{eee}
A_e(\rho_{\SC 1},\dots \rho_{\SC {4}})=\int dx \
K_{\rho_1}(x)K_{\rho_2}(x)K_{\rho_3}(x)K_{\rho_4}(x),
\end{equation} where we have dropped the $n$'s from our previous 
notation, since now they all take the value zero.
This expression is finite, and its explicit value is computed in
\cite{BC2}.  Finally, the vertex amplitude (\ref{choclo}) results  
\begin{eqnarray}
    \label{vv}
    \nonumber && A_v(\rho_{\SC v_1}, \dots
\rho_{\SC v_{10}}) = \int dx_1 \dots dx_5 \
K_{\rho_1}(x_1,x_5)K_{\rho_2}(x_1,x_4)K_{\rho_3}
(x_1,x_3)K_{\rho_4}(x_1,x_2)\\
&& \ \ \ \ \ \ \ \ \ \ \ \ \ \ \ \ \ \ \ \ K_{\rho_5}(x_2,x_5)
K_{\rho_6}(x_2,x_4)K_{\rho_7}(x_2,x_3)
K_{\rho_8}(x_3,x_5)K_{\rho_9}(x_3,x_4) K_{\rho_{10}}(x_4,x_5).
\end{eqnarray}
We can now remove the trivial divergence (the integration over the
gauge group) by dropping one of the group integrations (see footnote
\ref{foot} above).  The vertex amplitude (\ref{vv}) is precisely the
one defined by Barrett and Crane in \cite{BC2}.  The spin foam model
is finally given by
\begin{eqnarray}
    A(J)=\int_{\rho_f} d\rho_f \prod_{{f }}\ \rho_f^2 \ \
    \prod_{{e }} \ A_e( \rho_{\SC e_1},\dots \rho_{\SC e_{4}}) \
    \prod_{{ v }} \ A_v(\rho_{\SC v_1}, \dots \rho_{\SC v_{10}}),
    \end{eqnarray} 
It corresponds to the Lorentzian
generalization to the one defined in \cite{ac}.

\section{Discussion}

We have carried over the generalization of the model defined in
\cite{ac} to the Lorentzian signature.  The model is given by an
$SL(2,C)$ BF quantum theory plus a quantum implementation of the
additional constraints that reduce BF theory to Lorentzian general
relativity.

The analog model in the Euclidean $SO(4)$ case was shown to be finite
up to first bubble corrections.  It would be very interesting to study
this issue in the Lorentzian case.  Evidence in favor of the
conjecture of finiteness comes from the fact that, as
in the Euclidean case, the edge contribution in the model tends to
regularize the amplitudes.  Divergences appear when compatibility
conditions at edges fail to prevent colors associated to faces to get
arbitrarily large.  This happens when there are close surfaces in the
spin foam, namely, bubbles.  In \cite{ac} this divergences were cured by the
dumping effect of edge amplitudes.  As in its Euclidean relative, in
the Lorentzian model presented here the edge amplitude goes to zero
for large values of the colors.  More precisely, the amplitude
(\ref{eee}) behaves like $(\rho_1 \rho_2 \rho_3 \rho_4)^{-1}$ for
$\rho_i \rightarrow \infty$.

The state sum contains only representations of the form $(0,\rho)$. 
These correspond to the simple irreducible representations
representing space-like directions\cite{BC2}.  To obtain full general
relativity, it might be necessary to generalize the present
construction to include the others simple representations; that is,
those of the form $(n, 0)$, with $n$ an arbitrary integer, which
correspond to time-like directions.  A simple modification of the
action (\ref{action}) should allow these other balanced representation
to be included.  

These important issues will be investigated in the future.

\section{Acknowledgments}  

We thank Louis Crane for a very instructive discussion in a hot summer
day Marseillais.  This work was partially supported by NSF Grant
PHY-9900791. 

\begin{appendix}
      
\section{Representation Theory of $SL(2,C)$}

We review a series of relevant facts about $SL(2,C)$ representation 
theory. Most of the material of this section can be found in \cite{gel,ru}.
For a very nice presentation of the subject see also \cite{ted}.

We denote an element of $SL(2,C)$ by
\begin{equation}\label{g}
g=\left[\begin{array}{c} \alpha \ \beta \\
\gamma \ \delta  \end{array}\right],
\end{equation}
with $\alpha$, $\beta$, $\gamma$, $\delta$ complex numbers such that
$\alpha \delta-\beta \gamma=1$.  All the finite dimensional
irreducible representations of $SL(2,C)$ can be cast as a
representation over the set of polynomials of two complex variables
$z_1$ and $z_2$, of order $n_1-1$ in $z_1$ and $z_2$ and of order
$n_2-1$ in $\bar z_1$ and $\bar z_2$. The representation is given
by the following action
\begin{equation}
T(g)P(z_1,z_2)=P(\alpha z_1 +\gamma z_2,\beta z_1 +\delta z_2).
\end{equation}
The usual spinor representations can be directly related to these ones.
 
 The infinite dimensional representations are realized over the space
of homogeneous functions of two complex variables $z_1$ and $z_2$ in
the following way.  A function $f(z_1, z_2)$ is called homogeneous of
degree $(a,b)$, where $a$ and $b$ are complex numbers differing by an
integer, if for every $\lambda \in C$ we have
\begin{equation}\label{hom} f(\lambda z_1, \lambda z_2)=\lambda^a
\bar\lambda^b f(z_1, z_2),
\end{equation}
where $a$ and $b$ are required to differ by an integer in order to
$\lambda^a \bar\lambda^b$ be a singled valued function of $\lambda$. 
The infinite dimensional representations of $SL(2,C)$ are given by the
infinitely differentiable functions $f(z_1, z_2)$ (in $z_1$ and $z_2$
and their complex conjugates) homogeneous of degree $({\mu+n\over
2},{\mu-n \over 2})$, with $n$ an integer and $\mu$ a complex number. 
The representations are given by the following action
\begin{equation}\label{iii}
T_{n \mu}(g)f(z_1,z_2)=f(\alpha z_1 +\gamma z_2,\beta z_1 +\delta z_2).
\end{equation}
One simple realization of these functions is given by the functions of
one complex variables defined as \begin{equation} \phi(z)=f(z,1).
\end{equation}
On this set of functions the representation operators act in the
following way
\begin{equation}\label{actionn}
T_{n \mu}(g)\phi(z)=(\beta z +\delta)^{{\mu+n\over 2}-1}( {\bar \beta
\bar z +\bar \delta})^{{\mu-n\over 2}-1}\phi \left({\alpha z +\gamma
\over \beta z +\delta}\right).  
\end{equation} 
Two representations $T_{n_1\mu_2}(g)$ and $T_{n_1\mu_2}(g)$ are
equivalent if $n_1=-n_2$ and $\mu_1=-\mu_2$.

Unitary representations of $SL(2,C)$ are infinite dimensional.  They
are a subset of the previous ones corresponding to the two possible
cases: $\mu$ purely imaginary ($T_{n,i \rho}(g)$ $\mu=i\rho$,
$\rho=\bar \rho$, known as the {\em principal series}), and $n=0$,
$\mu=\bar \mu=\rho$, $\rho \neq 0$ and $-1 < \rho < 1$ ($T_{0
\rho}(g)$the {\em supplementary series}).  From now on we concentrate
on the principal series unitary representations $T_{n i\rho}(g)$ which
we denote simply as $T_{n \rho}(g)$ (dropping the $i$ in front of
$\rho$).  The invariant scalar product for the principal series is
given by \begin{equation} (\phi,\psi)= \int \bar\phi(z) \psi(z) dz,
\end{equation}
where $dz$ denotes $dRe(z) dIm(z)$.

There is a well defined measure on $SL(2,C)$ which is right-left
invariant and invariant under inversion (namely,
$dg=d(gg_0)=d(g_0g)=d(g^{-1})$).  Explicitly, in terms of the
components in (\ref{g})
\begin{eqnarray}\label{mea}
dg = \left( {i \over 2}\right)^3 {d\beta d\gamma d\delta \over
|\delta|^2} = \left( {i \over 2}\right)^3 {d\alpha d\gamma d\delta
\over |\gamma|^2} = \left( {i \over 2}\right)^3 {d\beta d\alpha
d\delta \over |\beta|^2} = \left( {i \over 2}\right)^3 {d\beta d\gamma
d\alpha \over |\alpha|^2},\end{eqnarray} where $d\alpha$, $d\beta$,
$d\gamma$, and $d\delta$ denote integration over the real and
imaginary part respectively.

Every square-integrable function, i.e, $f(g)$ such that
\begin{equation}
\int |f(g)|^2 dg \le \infty,
\end{equation} has a well defined Fourier transform defined as
\begin{equation}
F(n,\rho)=\int f(g) T_{n,\rho}(g) dg.
\end{equation}
This equation can be inverted to express $f(g)$ in terms of
$T_{n,\rho}(g)$.  This is known as the Plancherel theorem which
generalizes the Peter-Weyl theorem for finite dimensional unitary
irreducible representations of compact groups as $SU(2)$.  Namely,
every square-integrable function $f(g)$ can be written as
\begin{equation}\label{fu}
f(g)={1\over 8 \pi^4}\sum_n \int {\rm
Tr}[F(n,\rho)T_{n,\rho}(g^{-1})](n^2+\rho^2) d\rho, \end{equation}
where only components corresponding to the principal series are summed
over (not all unitary representations are needed)\footnote{If the
function $f(g)$ is infinitely differentiable of compact support then
it can be shown that $F(n,\rho)$ is an analytic function of $\rho$ and
an expansion similar to (\ref{fu}) can be written in terms of
non-unitary representations.}, and
\begin{equation}
{\rm Tr}[F(n,\rho)T_{n,\rho}(g^{-1})]=\int {\cal F}_{n\rho}(z_1,z_2)
{\cal T}_{n\rho}(z_2,z_1;g) dz_1dz_2.\end{equation} ${\cal
F}_{n\rho}(z_1,z_2)$, and ${\cal T}_{n\rho}(z_2,z_1;g)$ correspond to
the kernels of the Fourier transform and representation respectively
defined by their action on the space of functions $\phi(z)$ (they are
analogous to the momenta components and representation matrix elements
in the case of finite dimensional representations), namely
\begin{equation}
F(n,\rho)\phi(z):= \int f(g)T_{n\rho}(g)\phi(z) dg:=\int {\cal
F}_{n\rho}(z,\tilde z) \phi(\tilde z) d\tilde z, \end{equation} and
\begin{equation}
T_{n,\rho}(g)\phi(z):= \int {\cal
T}_{n\rho}(z,\tilde z;g) \phi(\tilde z) d\tilde z.\end{equation}
From (\ref{actionn}) we obtain that 
\begin{equation}
{\cal T}_{n\rho}(z,\tilde z;g)=(\beta z +\delta)^{{\rho+n\over 2}}( {\bar
\beta \bar z +\bar \delta})^{{\rho-n\over 2}} \delta\left(\tilde z-{\alpha z
+\gamma \over \beta z +\delta}\right).
\end{equation}
The ``resolution of the identity'' takes the form
\begin{equation}\label{fu1}
\delta(g)={1\over 8 \pi^4}\sum_n \int {\rm
Tr}[T_{n,\rho}(g)](n^2+\rho^2) \ d\rho.
\end{equation} This is a key formula that we use in the paper.

There exists an alternative realization of the representations in terms of
the space of homogeneous functions $f(z_1,z_2)$ defined above\cite{ru}.
Because of homogeneity (\ref{hom}) any $f(z_1,z_2)$ is completely determined
by its values on the sphere $S^3$
\begin{equation}\label{up}
|z_1|^2+|z_2|^2=1.
\end{equation}
As it is well now there is an isomorphism between $S^3$ and $SU(2)$ given by
\begin{equation}
u=\left[\begin{array}{c}\bar z_2 \ -\bar z_1 \\ z_1 \ \ \ \ \, z_2 
\end{array}\right] \end{equation} for $u\in SU(2)$ and $z_i$ satisfying
(\ref{up}). Alternatively we can define the the function $\phi(u)$ of $u\in
SU(2)$ as \begin{equation}
\phi(u):=f(u_{21},u_{22}),
\end{equation} with $f$ as in (\ref{hom}).
Due to (\ref{hom}) $\phi(u)$  has the following ``gauge'' behavior 
\begin{equation}\label{gauge}
\phi(\gamma u)= e^{i\omega(a-b)}\phi(u)= e^{i\omega n}\phi(u),
\end{equation}
for $\gamma= \left[ \begin{array}{c} e^{i \omega} \ 0 \\ 0 \ e^{-i
\omega} \end{array} \right]$.  The action of $T_{n \rho}(g)$ on
$\phi(u)$ is induced by its action on $f(z_1,z_2)$ (\ref{iii}).  We
can now use Peter-Weyl theorem to express $\phi(u)$ in terms of
irreducible representations $D^j_{q_1q_2}(u)$ of $SU(2)$; however in
doing that one notices that due to (\ref{gauge}) only the functions
$\phi^j_q(u)=(2j+1)^{\SC 1/2} D^j_{n q_2}(u)$ are needed (where
$j=|n|+k$, $k=0,1,\dots$).  Therefore $\phi(u)$ can be written as
\begin{equation}
\phi(u)=\sum^{\infty}_{j=n} \sum^j_{q=-j} d^j_q \ \ \phi^j_q(u).
\end{equation} 
This set of functions is known as the canonical basis.  This basis is
better suited for generalizing the Euclidean spin
foam models, since the notation maintains a certain degree of
similarity with the one in \cite{dfkr,ac}.  We can use this basis to
write the matrix elements of the operators $T_{n,\rho}(g)$, namely
\begin{equation}\label{ma}
D^{n \rho}_{j_1 q_1 j_2 q_2}(g)=\int_{\SC SU(2)} \bar
\phi^{j_1}_{q_1}(u) \ \left[T_{n \rho}(g) \phi^{j_2}_{q_2}(u)\right]
du.  \end{equation} Since $T_{n_1 n_2}(u_0)\phi(u)=\phi(u_0 u)$,
invariance of the $SU(2)$ Haar measure implies that
\begin{equation}\label{su}
D^{n \rho}_{j_1 q_1 j_2 q_2}(u_0)=\delta_{j_1 j_2} \ D^{j_1}_{q_1 q_2}(u_0).
\end{equation} In terms of these matrix elements equation (\ref{fu})
acquires the more familiar form
\begin{equation}
f(g)=\sum^{\infty}_{n=0} \int^{\infty}_{\rho=0}
\left[\sum^{\infty}_{j_1,j_2=n} \sum^{j_1}_{q_1=-j_1}
\sum^{j_2}_{q_2=-j_2} \bar D^{n,\rho}_{j_1 q_1 j_2 q_2}(g) f^{j_1 q_1
j_2 q_2}_{n,\rho} \right](n^2+\rho^2) d\rho, \end{equation} where
\begin{equation}
f^{j_1 q_1 j_2 q_2}_{n,\rho}=\int f(g) D^{n,\rho}_{j_1 q_1 j_2 q_2}(g) dg,
\end{equation}
and the quantity in brackets
represents the trace in (\ref{fu}). In the same way we can translate equation
(\ref{fu1}) obtaining
\begin{equation}\label{vani}
\delta(g)=\sum^{\infty}_{n=0} \int^{\infty}_{\rho=0}
\left[\sum^{\infty}_{j=n} \sum^{j}_{q=-j} \bar D^{n,\rho}_{j q j q}(g) 
\right](n^2+\rho^2) d\rho.
\end{equation}  

Using equations (\ref{ma}) and (\ref{su}), we can compute
\begin{eqnarray}\label{proj}
\int_{\SC SU(2)}\! \! \!  D^{n,\rho}_{j_1 q_1 j_2 q_2}(u)\ du = 
\delta_{j j_2}\int_{\SC
SU(2)} D^{j}_{q  q_2}(u) du 
 =  \delta_{j_2 0} \delta_{j_1 0},
\end{eqnarray} 
a second key equation for the paper.

\end{appendix}

\end{document}